\documentclass[12pt]{iopart}

\usepackage{booktabs}
\usepackage{graphicx}
\usepackage{subfigure}
\usepackage{indentfirst}
\usepackage{bm}
\usepackage{multirow}

\begin{document}

\title[]{The effect of single-particle space-momentum angle distribution on two-pion HBT correlation in high-energy heavy-ion collisions}

\author{Hang Yang, Qichun Feng, Yanyu Ren, Jingbo Zhang*, Lei Huo}

\address{School of Physics, Harbin Institute of Technology, Harbin 150001, China}

\ead{jinux@hit.edu.cn}
\vspace{10pt}

\begin{abstract}
Using several source models, we analyze the transverse momentum dependence of HBT radii in the relativistic heavy-ion collisions. The results indicate that the single-particle space-momentum angle distribution plays an important role in the transverse momentum dependence of HBT radii. In a cylinder source, we use several formulas to describe the transverse momentum dependence of HBT radii and the single-particle space-momentum angle distribution. We also make a numerical connection between them in the transverse plane. 
\end{abstract}
\vspace{2pc}
\maketitle
\section{Introduction}
A new state of matter has been found in the Relativistic Heavy Ion Collider, which is called Quark-Gluon-Plasma (QGP)\cite{CABIBBO197567,BACK200528,SHI2009187c}. It is a strongly interacting partonic matter formed by deconfined quarks and gluons under extreme temperature and energy density. This state is similar to the early time of the universe after the big bang\cite{SATZ2001204}, and it has aroused great interest in the physics community. A powerful tool in studying the mechanism of particle production in hot QCD matter is the two-pion intensity interferometry. The interferometry analyses were first shown by Hanbury Brown and Twiss to measure the angular diameter of stars in the 1950s, and obtained the name, the HBT method. Then G. Goldhaber, S. Goldhaber, W. Lee and A. Pais extended this method in $\overline{p}+p$ collisions\cite{PhysRev.120.300}. After that, the two-pion interferometry has been widely used in high-energy heavy-ion collisions, with much development and improvement. For example, the HBT radii parameters may be used to located the Critical End Point (CEP) in the QCD phase diagram\cite{PhysRevLett.114.142301}, and the multi-pion interferometry has been used in high-energy heavy-ion collisions, as an extension of two-pion interferometry\cite{Biyajima:1980qi,PhysRevC.34.1667,Bary_2018}. 

\begin{figure}[htb]
	\centering
	\includegraphics[scale=0.125]{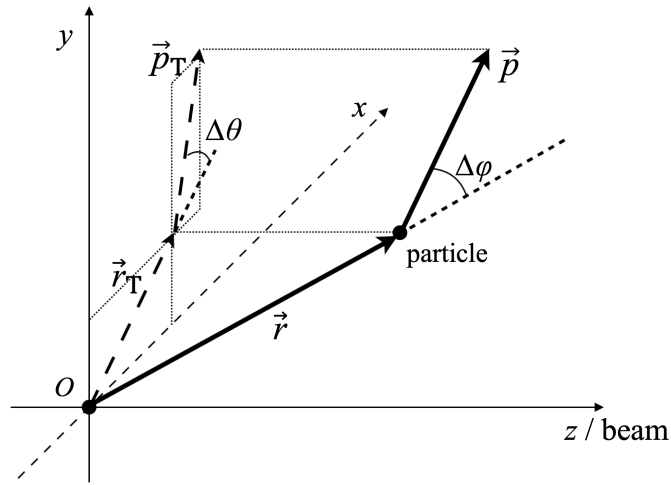}
	\caption{The diagram of the $\Delta\varphi$ and $\Delta \theta$. $\Delta\varphi$ is the angle between $\vec r$ and $\vec p$, and $\Delta \theta$ is the angle between $\vec r_{\rm T}$ and $\vec p_{\rm T}$, at the freeze out time. The origin is the center of the source.}
	\label{figpr}
\end{figure}

Many collaborations use the HBT method to analyze different collisions at different energies\cite{2011328,PhysRevC.71.044906,Kniege_2004,PhysRevD.84.112004,PhysRevC.92.014904}. Most of them show the phenomenon of transverse momentum or transverse mass dependence of HBT radii. And the shrinking of HBT radii with increasing transverse momentum is associated with collective flow\cite{doi:10.1146/annurev.nucl.55.090704.151533,Lisa:2008gf}. At the energy range of the Beam Energy Scan Phase II(BES-II) at RHIC, the flow is not so strong compared with the flow at the LHC energy range\cite{Ghosh_2014}, when the particles freeze out, there will be a finite angle between the radius vector and the momentum vector. We named it as the single-particle space-momentum angle $\Delta\varphi$. Figure~\ref{figpr} is the diagram of this angle and its projection angle $\Delta \theta$ on the transverse plane.

We use locally thermalized fireballs with the collective flow to produce the particles, which are the same as the blast wave model\cite{PhysRevLett.42.880}. The blast wave model has already been used in analyses of the HBT correlations\cite{PhysRevC.70.044907,PhysRevC.73.064902,PhysRevC.68.034913}, and it can also be used to describe the transverse momentum dependence on HBT radii\cite{Zhang:2016tbf,Bialas_2015}. The space-momentum correlation has a great influence on the HBT radii. So in this paper, we use the single-particle space-momentum angle distribution to describe the space-momentum correlation. And we focus on the $\Delta\varphi$ distribution effect on the transverse momentum dependence of HBT radii. Then we try to build a new connection between the $\Delta \theta$ distribution and the transverse momentum dependence of HBT radii, in the transverse plane. By this connection, we can use the single-particle space-momentum angle distribution to describe the transverse momentum dependence of HBT radii, in the transverse plane.

This paper is structured as follows. Sec. 2 briefly introduces the CRAB code and the method used to calculate the HBT radii. In Sec. 3, we calculate the HBT radii for pions in different sources. In Sec. 4, a numerical connection has been built between the $\Delta \theta$ angle distribution and the transverse momentum dependence of $R_{\rm o}$, $R_{\rm s}$. Finally, we summarize our conclusions in Sec. 5.

\section{CRAB code and methodology}
In this paper, we use the Correlation After Burner (CRAB) code to read the phase-space information of generated pions and calculate the two-pion correlation functions\cite{CRAB:2006}. The code is based on the formula
\begin{equation}
	C(\bm{q},\bm{K})=1+\frac{\int {\rm d}^4x_1 {\rm d}^4x_2 S_1(x_1,{\bm p}_2) S_2(x_2,{\bm p}_2)
		{\left|\psi_{\rm{rel}} \right|}^{2}}
		{\int {\rm d}^{4}x_{1}{\rm d}^{4}x_{2}S_{1}(x_{1},{\bm p}_{2})S_{2}(x_{2},{\bm p}_{2})},
\end{equation}
where ${\bm q}={\bm p}_1-{\bm p}_2$, ${\bm K}=({\bm p}_1+{\bm p}_2)/2$, and $\psi_{\rm{rel}}$ is the two particle wave function. In further discussion, we neglect the Coulomb interaction and strong interactions between pions. The correlation functions can be calculated in different $p_{\rm T}$ bins by changing the kinematic cuts in the fitter of the CRAB code. We use the single-pion information extracted from the calculation to analyze the space-momentum angle $\Delta\varphi$ distribution.  

\begin{figure}[htb]
	\centering
	\includegraphics[scale=0.25]{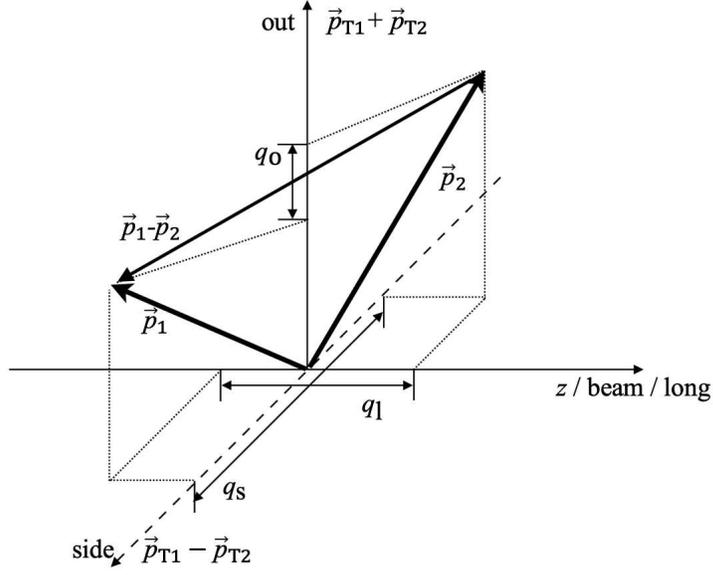}
	\caption{The diagram of `out-side-long'(o-s-l) coordinate system.}
	\label{figo-s-l}
\end{figure}

We usually use the `out-side-long'(o-s-l) coordinate system in the HBT research, shown in Figure~\ref{figo-s-l}. The longitudinal direction is along the beam direction, and the transverse plane is perpendicular to the longitudinal direction. In the transverse plane, the momentum direction of pair particles is the outward direction. And the direction perpendicular to the outward direction is called the sideward direction.

\begin{figure}[htb]
	\centering
	\includegraphics[scale=0.15]{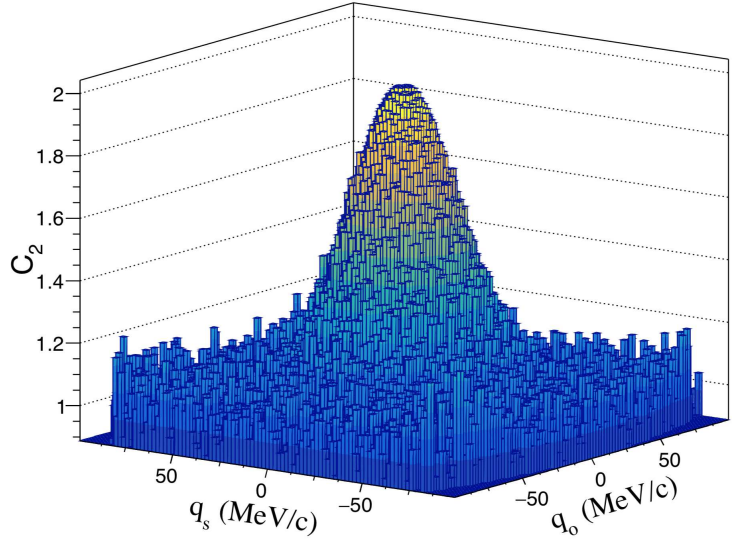}
	\caption{HBT correlation function in $q_{\rm o}$ and $q_{\rm s}$ directions for a Gaussian source.}
	\label{correlation_function}
\end{figure}

In this paper, when calculating the HBT correlation function, the rapidity range is always set to $-0.5<\eta<0.5$. An example of correlation functions of a Gaussian source is shown in Figure~\ref{correlation_function}. It is in $q_{\rm o}$ and $q_{\rm s}$ directions, with $-3<q_{\rm l}<3$ MeV/c.

The HBT correlation function of the Gaussian form can be written as\cite{Alexander_2003}
\begin{equation}
\label{fit_function}
	C(\bm{q},\bm{K})=1+\lambda {\exp}
		{[-q_{\rm o}^2 R_{\rm o}^2(\bm K)-q_{\rm s}^2 R_{\rm s}^2(\bm K)-
		q_{\rm l}^2 R_{\rm l}^2(\bm K)]},
\end{equation}
where $\lambda$ is coherence parameter. The $R_{\rm i}^2$ can be expressed as\cite{doi:10.1146/annurev.nucl.49.1.529,PhysRevC.53.918}
\begin{eqnarray}\label{r_side}
	R^2_{\rm s} = \langle r^2_{\rm s} \rangle,\\
	R^2_{\rm o} = \langle (r_{\rm o} -\beta _{\rm o} t)^2 \rangle -\langle r_{\rm o} -\beta _{\rm o} t \rangle^2,\\
	R^2_{\rm l} = \langle (r_{\rm l} -\beta _{\rm l} t)^2 \rangle -\langle r_{\rm l} -\beta _{\rm l} t \rangle^2,
\end{eqnarray}
where the average notation is defined as
\begin{equation}
	\langle \xi \rangle=\frac{\int d^4x \xi S(x,p)}{\int d^4x S(x,p)}.
\end{equation}
We can calculate the HBT radii by using equation~(\ref{fit_function}) to fit the HBT correlation function which is generated from the CRAB code.
\section{Transverse momentum dependence of HBT radii}

The single-particle momentum space angle distribution, also called $\Delta\varphi$ angle distribution, can directly cause the transverse momentum $p_{\rm T }$ dependence.

\begin{figure}[!hbt]
	\centering
	\subfigure[]
	{
		\includegraphics[width=0.4\textwidth]{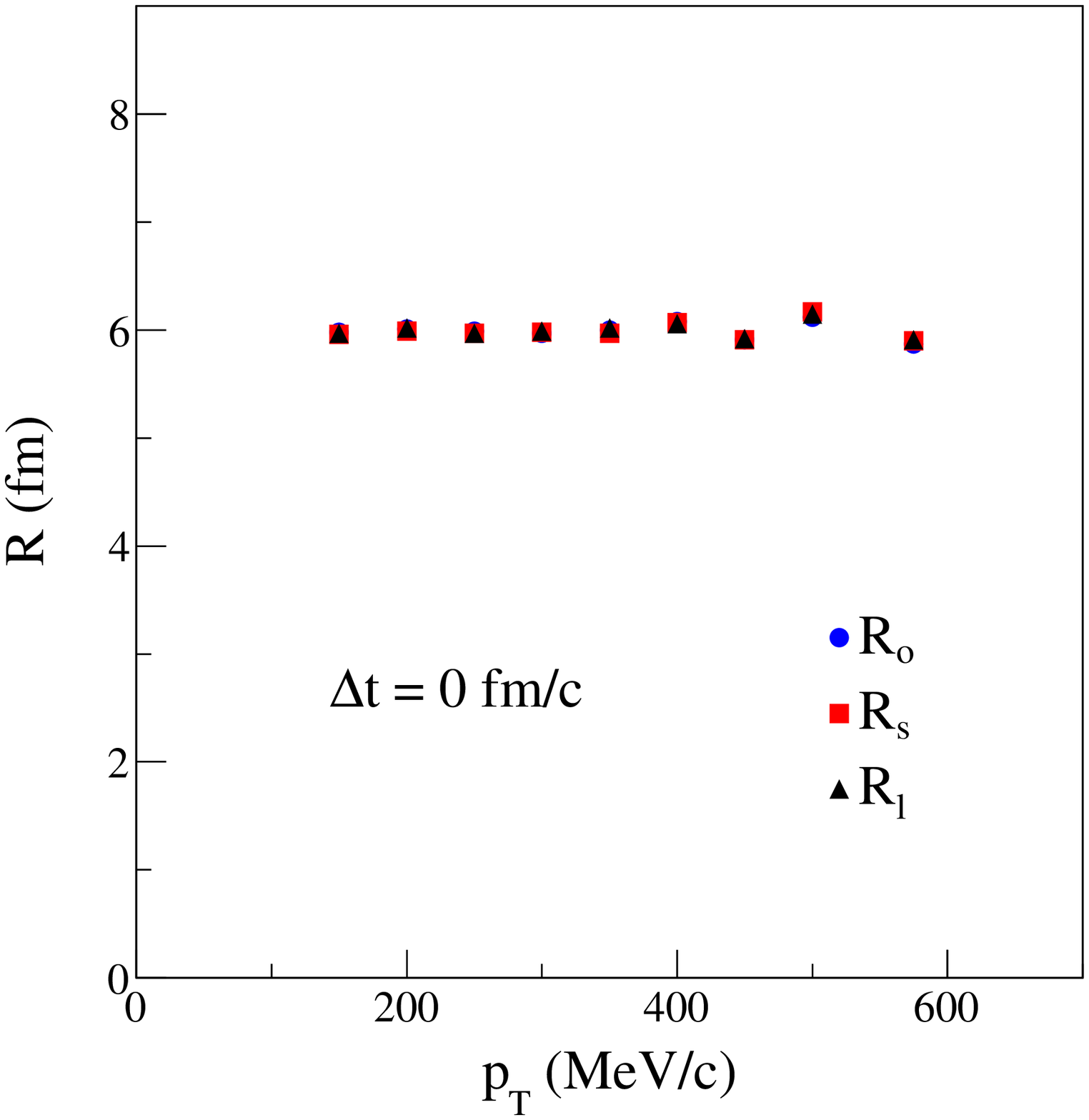}
	}
	\subfigure[]
	{
		\includegraphics[width=0.4\textwidth]{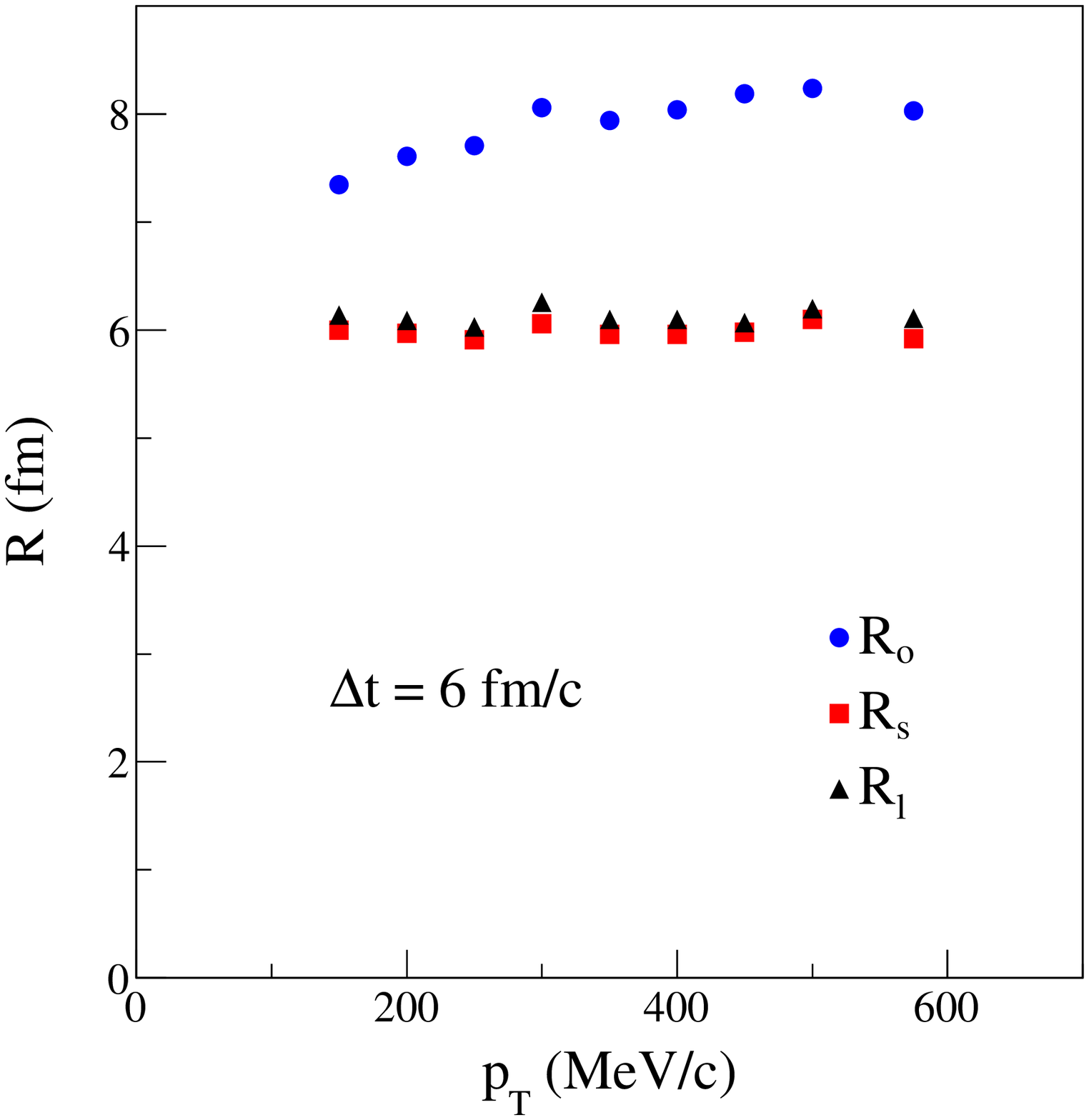}
	}
	\caption{Transverse momentum dependence of HBT radii for a Gaussian source.}
	\label{gaussian_source}
\end{figure}

Firstly, the influence of source lifetime need be excluded. With a Gaussian source to generate pion data, the emission function can be written as
\begin{equation}\label{Gaussian}
	S(x,{\bm p})=A\bm p^{2}{\rm{exp}}\left(-\frac{\sqrt{\bm p^2+m^2}}{T}\right) 
		{\rm{exp}}\left(-\frac{\bm r^2}{2R^2}-\frac{t^2}{2(\Delta t)^2}\right),
\end{equation}
where we always set source size $R=6.0$ fm, temperature $T=100$ MeV, and mass of pions $m=139.58$ $\rm MeV/c^2$. Then we use the CRAB code to calculate the HBT correlation functions of pions.  After that, we use equation~(\ref{fit_function}) to fit the correlation functions in different $p_{\rm T}$ bins, and there are 9 bins in 125 MeV/c $<p_{\rm T}<$ 625 MeV/c. The transverse momentum dependence of HBT radii are shown in Figure~\ref{gaussian_source}. 

\begin{figure}[htb]
	\centering
	\includegraphics[scale=0.4]{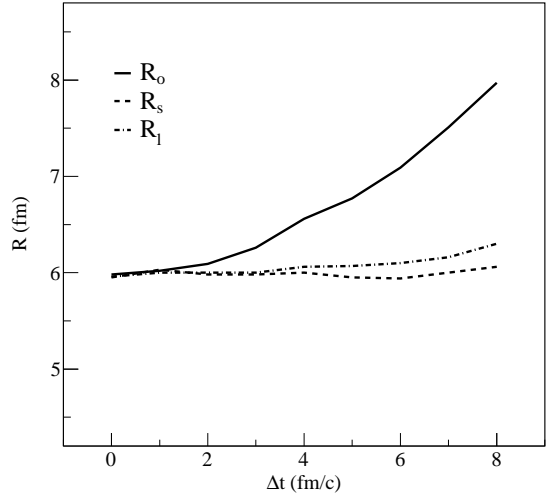}
	\caption{HBT radii changes with $\Delta t$ for a Gaussian source.}
	\label{dt_dependence}
\end{figure}

In Figure~\ref{gaussian_source}(a), when the lifetime of source is $\Delta t=0$ fm/c, all the pions freeze out at the same time. The HBT radii coincide with each other, and they are almost equal to the source radii. There is no transverse momentum dependence of HBT radii. With $\Delta t=6$ fm/c, as shown in Figure~\ref{gaussian_source}(b), the value of $R_{\rm{o}}$ increases gradually with the $p_{\rm T}$. There is little changes of $R_{\rm{l}}$ and $R_{\rm{s}}$. Then we calculate the HBT radii by changing the value of $\Delta t$ in Figure~\ref{dt_dependence}.

In Figure~\ref{dt_dependence}, the Gaussian source radius is still set to 6 fm, and the transverse momentum range is 125-625 MeV/c. We can see the increase of $R_{\rm o}$ at higher lifetime $\Delta t$ of source. Since the rapidity cut is $-0.5<\eta<0.5$, $R_{\rm l}$ only changes a little. There is barely any change in $R_{\rm s}$. We have already known the emission function of the Gaussian source, by using equation (\ref{r_side})-(\ref{Gaussian}), the HBT radius can be expressed as

\begin{eqnarray}
	R^2_{\rm s} = \langle r^2_{\rm s} \rangle,\\
	R^2_{\rm o} = \langle r^2_{\rm o} \rangle + {\langle \beta_{\rm o} \rangle}^2 \langle(\Delta t)^2 \rangle,\\
	R^2_{\rm l} = \langle r^2_{\rm l} \rangle+ {\langle \beta_{\rm l} \rangle}^2 \langle(\Delta t)^2 \rangle,
\end{eqnarray}
where, $r$ and $\beta$ are space coordinate and velocity of a single particle, respectively. Therefore we can reduce the influence of the lifetime of source by minimizing $\Delta t$.

For discussing the influence of single-particle angle distribution on the transverse momentum dependence of HBT radii, we introduce another source which is called space-momentum angle correlation source. the emission function can be written as 

\begin{equation}\label{space_momentum}
	S(x,{\bm p})=A\bm p^{2}{\rm{exp}}\left(-\frac{\sqrt{\bm p^2+m^2}}{T}\right) 
		{\rm{exp}}\left(-\frac{\bm r^2}{2R^2}-\frac{t^2}{2(\Delta t)^2}\right)
		{\rm{w}}\left(\Delta \varphi\right),
\end{equation}
where $\Delta \varphi$ is the single-pion space-momentum angle at freeze-out time. By changing the formula of ${\rm{w}}\left(\Delta \varphi\right)$, we can change the angle $\Delta \varphi$ distribution. If ${\rm{w}}\left(\Delta \varphi\right)=1$, the $\Delta \varphi$ value is totally random between $0-\pi$, and the source becomes a Gaussian source. Here, the function $\rm w$ in equation~(\ref{space_momentum}) can be written as
\begin{equation}
\label{cases}
	{\rm{w}}\left(\Delta \varphi \right)=\cases{0 & $\alpha < \Delta \varphi \leq \pi$\\ 
		1 & $0 \leq \Delta \varphi \leq \alpha$},
\end{equation}
where $\alpha$ is a given value. This function means that, only the pions whose angle $\Delta \varphi$ is smaller than the $\alpha$ value can exist. The radius of source is $R=6$ fm, and the lifetime is $\Delta t=0$ fm/c.
\begin{figure}[htb]
	\centering
	\subfigure[]
	{
		\label{alpha=0} 
		\includegraphics[width=0.4\textwidth]
		{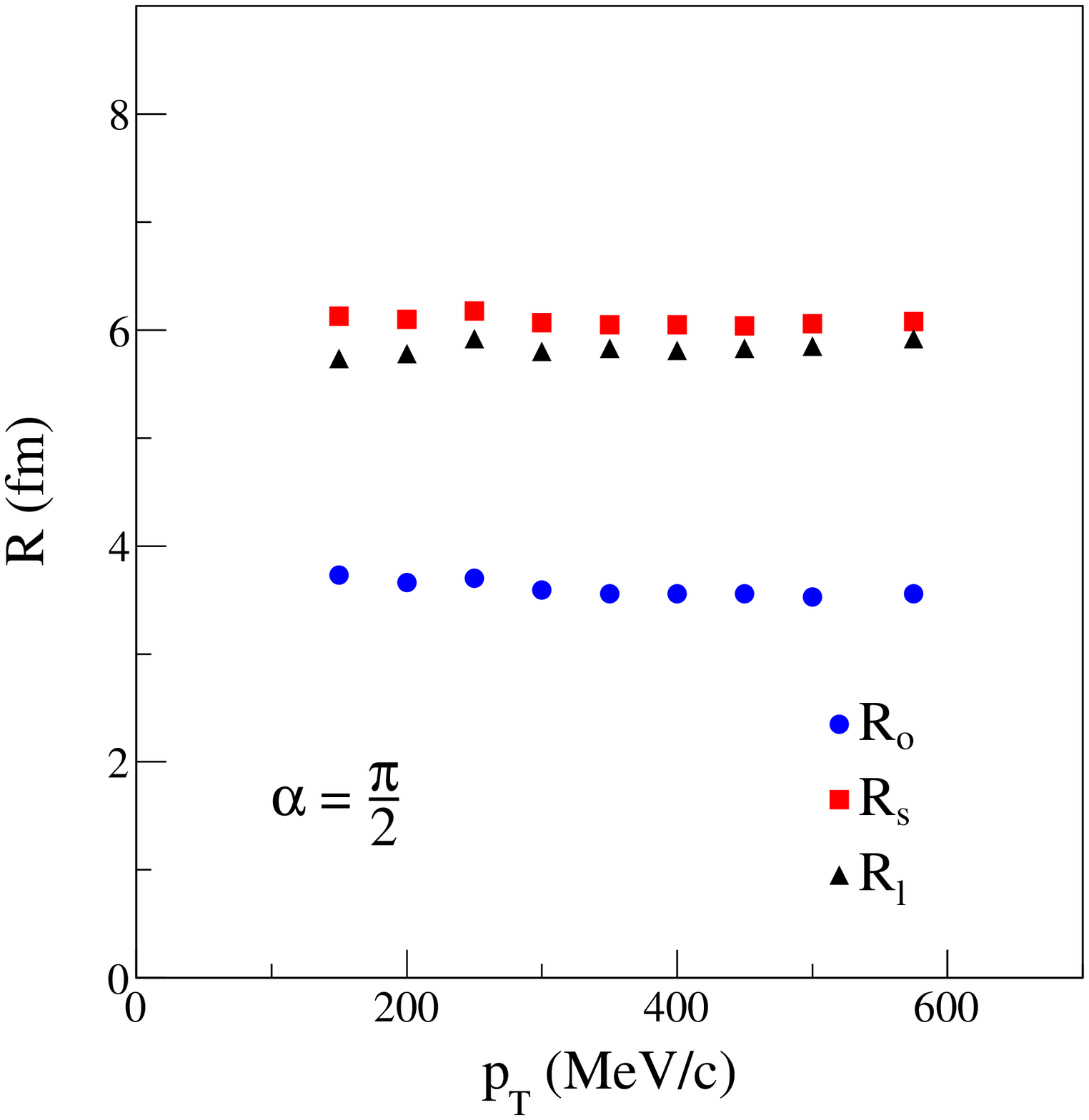}
	}
	\hspace{0.1cm}
	\subfigure[]
	{
		\label{alpha=0.5} 
		\includegraphics[width=0.4\textwidth]
		{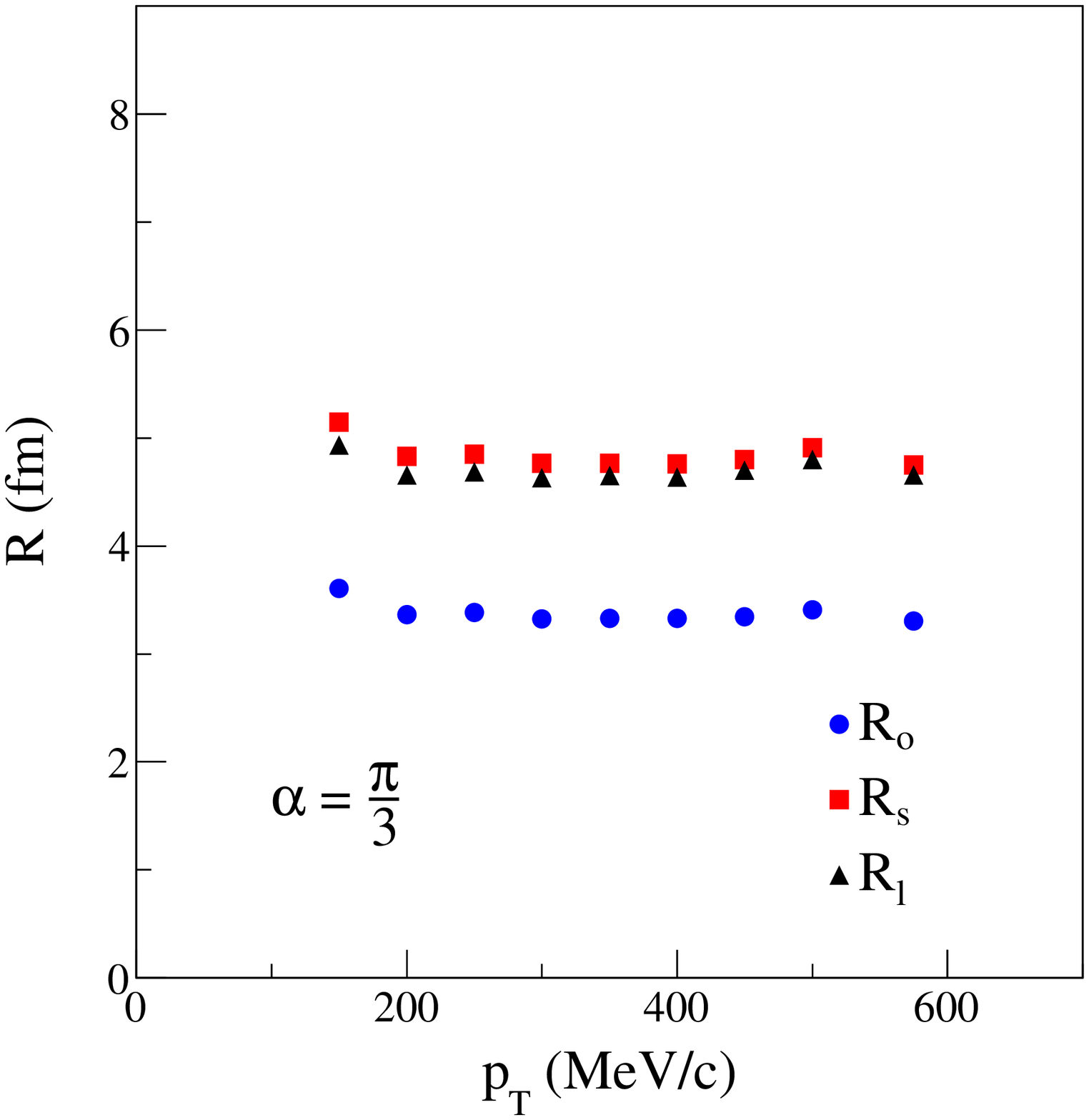}
	}
	\caption{HBT radii for a space-momentum angle correlation source.}
	\label{p_r_source}
\end{figure}

\begin{figure}[hbt]
	\centering
	\includegraphics[scale=0.6]{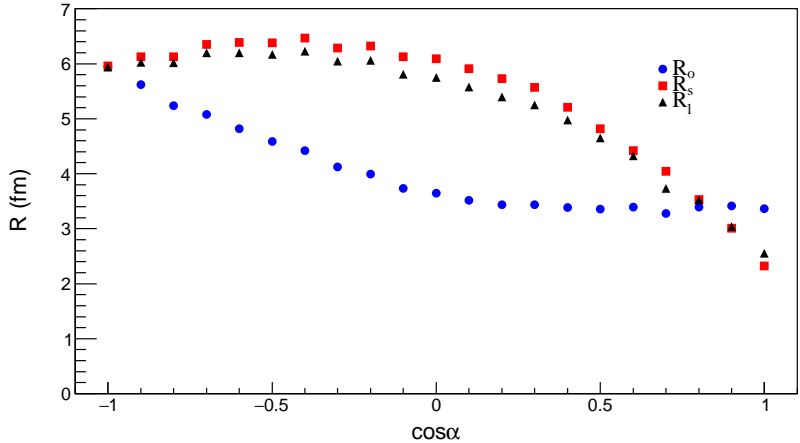}
	\caption[fig5]{$\cos \alpha$ dependence of HBT radii for a space-momentum angle correlation source.}
	\label{p_r_source_costheta}
\end{figure}

In Figure~\ref{p_r_source}(a), $\alpha=\frac{\pi}{2}$. $R_{\rm{o}}$ values are lower than $R_{\rm{s}}$ and $R_{\rm{l}}$, and the HBT radii are mostly flat as a function of $p_{\rm T}$. There is barely any appearance of $p_{\rm T}$ dependence. In Figure~\ref{p_r_source}(b), $\alpha=\frac{\pi}{3}$. Compared with the Figure~\ref{p_r_source}(a), the values of $R_{\rm{s}}$ and $R_{\rm{l}}$ becomes smaller. The values of $R_{\rm{o}}$ changes slightly. The present work shows that, the $\Delta\varphi$ distribution can affect the values of HBT radii. Furthermore, we still use the space-momentum angle correlation source to generate the data, the emission function and the w function are still based on the equation (\ref{space_momentum}) and equation (\ref{cases}), while the $\alpha$ varies between 0 and $\pi$. The HBT radii as the function of $\cos\alpha$ value is shown in Figure~\ref{p_r_source_costheta}.

In Figure~\ref{p_r_source_costheta}, the transverse momentum range is 125-625 MeV/c. When $-1<\cos\alpha<0$, the $R_{\rm{o}}$ values decrease with increasing $\cos\alpha$, while $R_{\rm{s}}$ and $R_{\rm{l}}$ only have small changes. When $0<\cos\alpha<1$, there is almost no changes of $R_{\rm{o}}$, while $R_{\rm{s}}$ and $R_{\rm{l}}$ decreasing with increasing $\cos\alpha$. The different space-momentum angle $\Delta\varphi$ distributions correspond to different HBT radii. Therefore, if we can control the $\Delta\varphi$ angle distribution for a given $p_{\rm T}$, we can reproduce the $p_{\rm T}$ dependence phenomenon.

\begin{figure}[htb]
	\centering
	\includegraphics[scale=0.5]{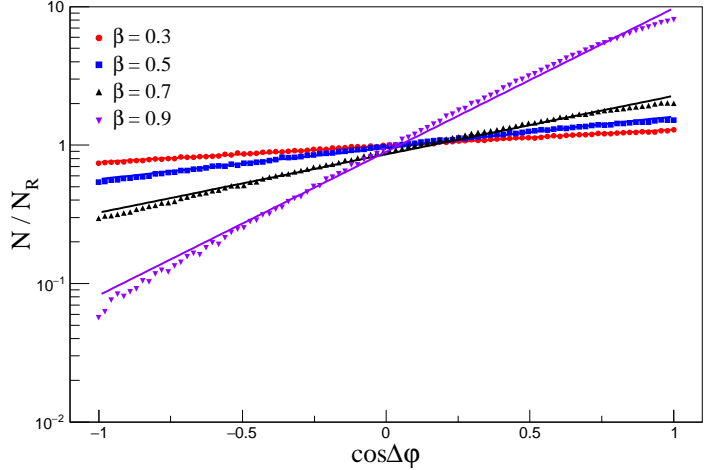}
	\caption{Fit normalized $\cos(\Delta\varphi)$ for different homogeneous expansion sources. The transverse momentum range is 125-175 MeV/c. The dots are normalized numbers of pions, and the lines are the fit lines.}
	\label{fit_p_r_angle}
\end{figure}

We use a homogeneous expansion source to calculate the HBT radii in different $p_{\rm T}$ regions, and then we use space-momentum angle correlation source to reproduce this phenomenon. The homogeneous expansion source is based on the Gaussian source. Every pion has been given an expansion velocity $\bm \beta$ along the $\bm r$ direction. And the momentum is generated by using Lorentz transformation. The emission function can be written as
\begin{equation}
	S=A\bm p^{2}{\rm{exp}}\left(-\frac{\gamma E-\gamma\bm {\beta p}}{T}\right) 
		{\rm{exp}}\left(-\frac{\bm r^2}{2R^2}-\frac{t^2}{2(\Delta t)^2}\right),
\end{equation}
where $\gamma=1/\sqrt{1-\beta^2}$ is the Lorentz factor, still with $R=6$ fm and the $\Delta t=0$ fm/c. After using this emission function to generate data, we use the CRAB code to calculate correlations and use equation(\ref{fit_function}) to calculate the HBT radii in different $p_{\rm T}$ regions. Meanwhile, we can also get the phase-space information of pions in different $p_{\rm T}$ regions. The equation to fit the normalized space-momentum angle distribution can be written as 
\begin{equation} \label{consinfit}
	f(\Delta\varphi)=c_1{\rm{exp}}(c_2\cos(\Delta\varphi)),
\end{equation} 
where, $c_1$ and $c_2$ are fit parameters. The fit lines are shown in Figure~\ref{fit_p_r_angle}.

\begin{figure}[hbt]
	\centering
	\subfigure[]
	{
		\label{simulate_expansion_u=0.3} 
		\includegraphics[width=0.4\textwidth,natwidth=600,natheight=715]
		{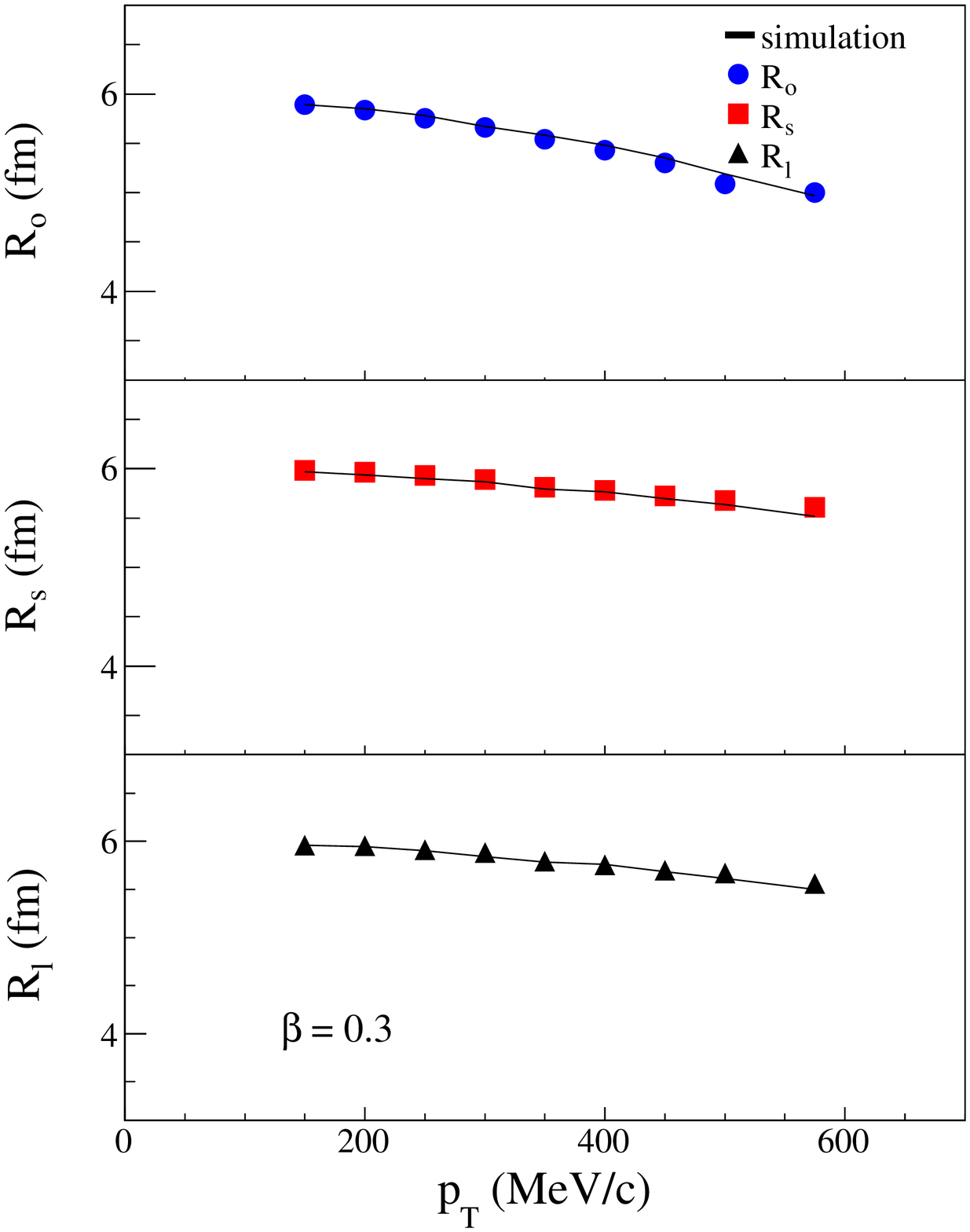}
	}
	\subfigure[]
	{
		\label{simulate_expansion_u=0.5}
		\includegraphics[width=0.4\textwidth,natwidth=600,natheight=715]
		{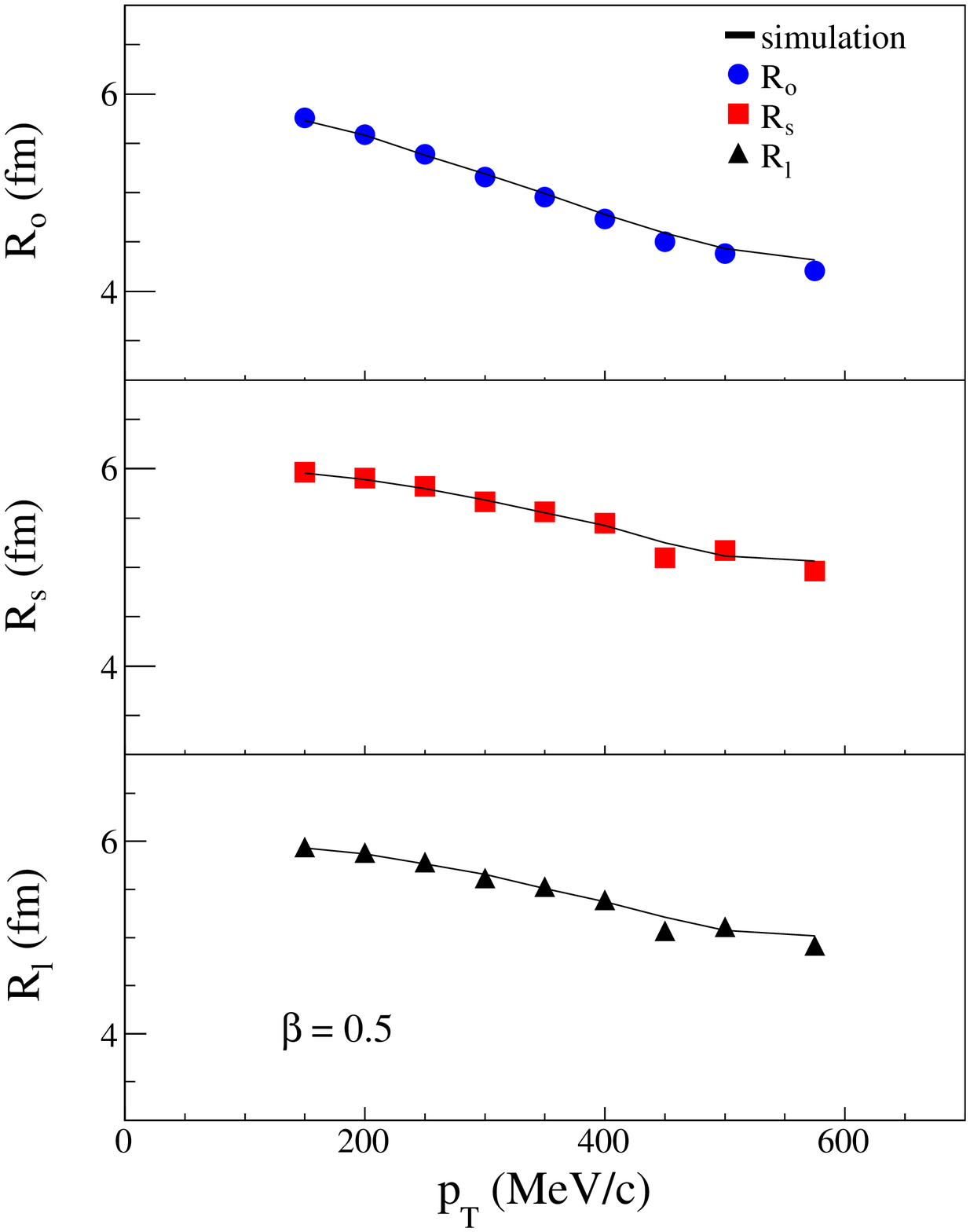}
	}
	\subfigure[]
	{
		\label{simulate_expansion_u=0.7} 
		\includegraphics[width=0.4\textwidth,natwidth=600,natheight=715]
		{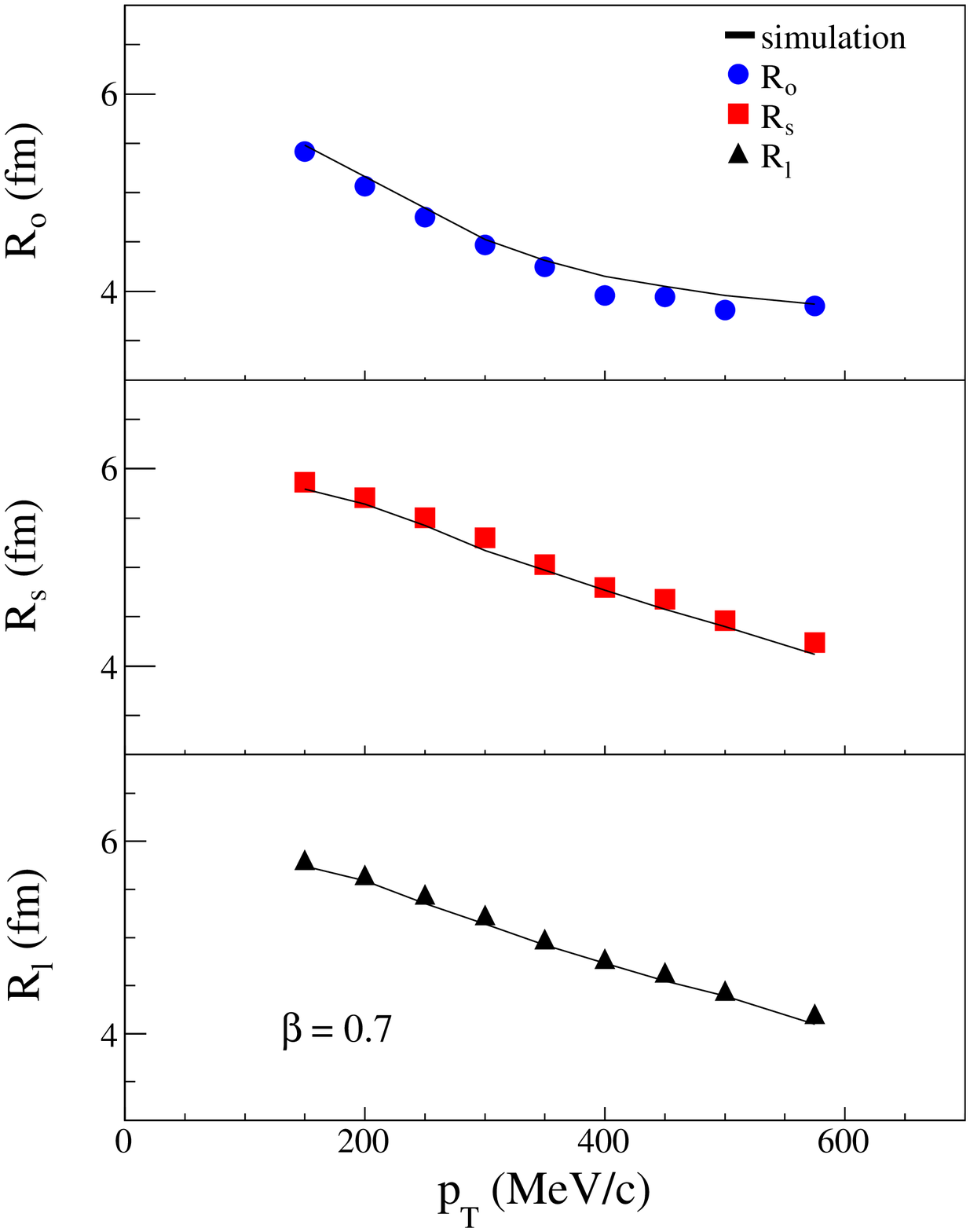}
	}
	\subfigure[]
	{
		\label{simulate_expansion_u=0.9}
		\includegraphics[width=0.4\textwidth,natwidth=600,natheight=715]
		{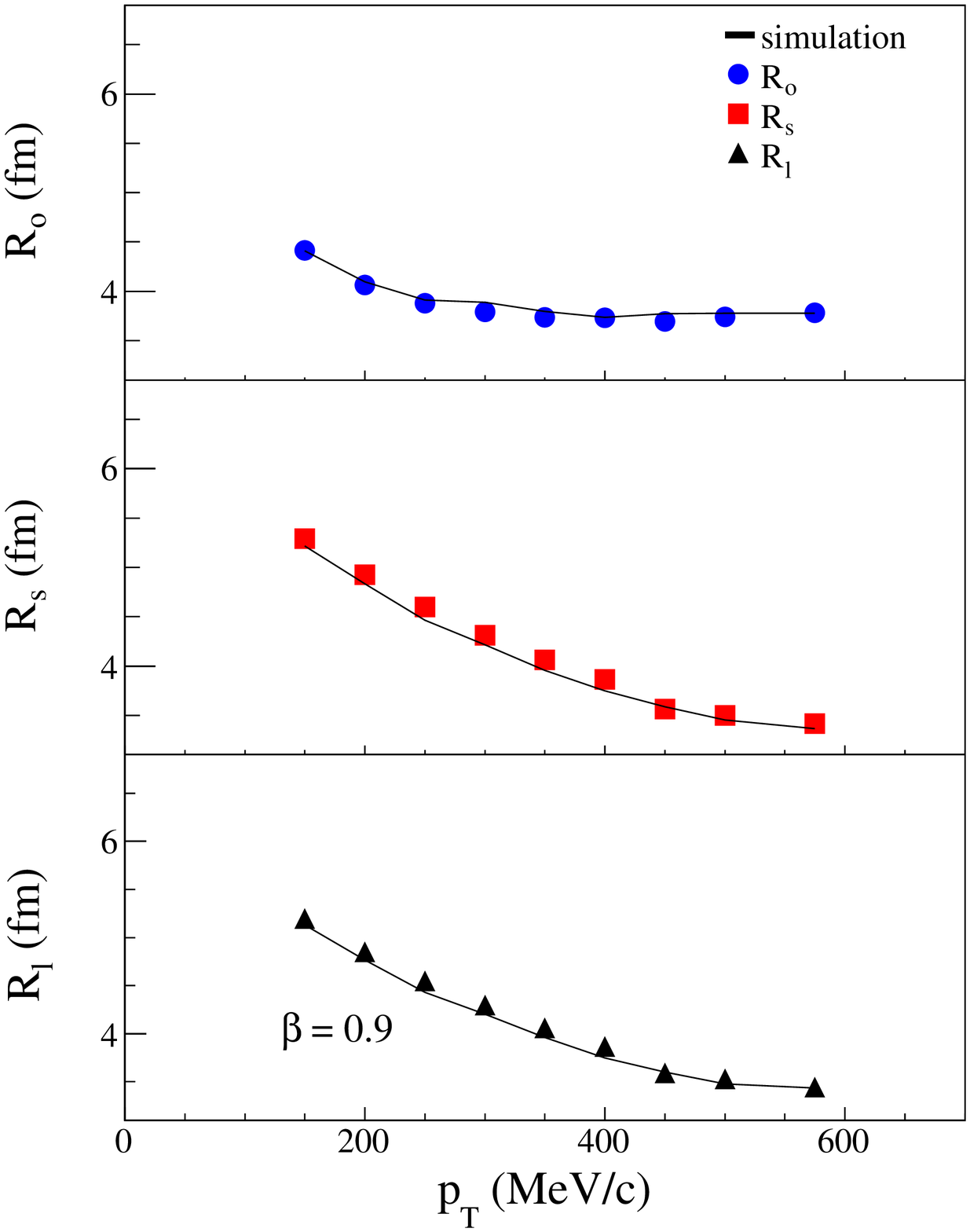}
	}
	\caption{The simulation for a homogeneous expansion source. The black simulation lines are calculated in space-momentum angle correlation source.}
	\label{simulation}
\end{figure}

There are 9 bins of $p_{\rm T}$, and we have 9 groups of two parameters $c_1$ and $c_2$. Then we use the space-momentum angle correlation source to simulate the HBT radii. In emission function (\ref{space_momentum}), ${\rm{w}}(\Delta\varphi)=f(\Delta\varphi)$. We put the values of $c_1$ and $c_2$ into this equation, and each $p_{\rm T}$ region corresponds to one emission function, so there are 9 emission functions. Then we use each emission function to generate the data and calculate the HBT radii in the corresponding $p_{\rm T}$ region, and both particles are taken from the same $p_{\rm T}$ region. After the calculation, the simulation results and the HBT radii calculated by homogeneous expansion source are shown in Figure~\ref{simulation}. 

For all the four situations in Figure~\ref{simulation}, the simulated HBT radii are almost equal to the HBT radii calculated in the homogeneous expansion sources. It indicates that the source expansion can cause the space-momentum angle $\Delta\varphi$ distribution narrowing down with transverse momentum. This meets our expectation that the different space-momentum angle $\Delta\varphi$ distributions lead to different values of HBT radii, which will then cause the transverse momentum dependence of HBT radii.

\section{Space-momentum angle distribution in transverse plane}

We use cylinder expansion source\cite{PhysRevC.53.918} to get the connection between the HBT radii and $\Delta \theta$(angle between $\vec p_{\rm T}$ and $\vec r_{\rm T}$) distribution. It can be written as
\begin{eqnarray}
	\fl S\left(x,p\right)= A{M_{\rm T} {\cosh}\left( \eta-Y \right)}
		{\rm{exp}}\left(-\frac{pu\left(x\right)}{T}\right)
		{\exp}\left(-\frac{\left( \tau-\tau_{0} \right)^2}{2\left(\delta\tau\right)^2}
		-\frac{\rho^2}{2R^2_{\rm g}} - \frac{\eta^2}{2\left( \delta \eta \right)^2}\right),
\end{eqnarray}
where $u(x)$ is the 4-velocity and can be decomposed as 
\begin{equation}
	u(x)=\left( \cosh\eta\cosh\eta_{\rm T},\sinh\eta_{\rm T}\vec{\rm e}_{\rm T},\sinh\eta\cosh\eta_{\rm T} \right), 
\end{equation}
and $\eta=\frac{1}{2}\ln[(p+z)/(p-z)]$ is the longitudinal flow rapidities. The transverse flow rapidity is defined as 
\begin{equation}
	\eta_{\rm T}=\cases{\eta_{\rm Tmax}\frac{\rho}{R_{\rm g}} & $\rho < R_{\rm g}$\\ 
		\eta_{\rm Tmax} & $\rho \geq R_{\rm g}$}.
\end{equation}
The rapidity of the pion is $Y=\frac{1}{2}\ln[(E+p_{\rm l})/(E-p_{\rm l})]$, the proper time is $\tau=\sqrt{t^2-z^2}$, and the pion radial position in the transverse plane is $\rho = \sqrt{x^2+y^2}$. We set $T = 100$ MeV, $\delta\tau = 0$ fm/c, $\tau_0$ = 10 fm/c, $R_{\rm {g}}$ = 6.0 fm and $\delta\eta$ = 3.0, and the variable is $\eta_{\rm Tmax}$.

Since the CRAB filter is set $-0.5<\eta<0.5$ and $\delta\tau = 0$ fm/c, all pions almost freeze out at the same time ($\Delta t<1.3$fm/c). The effect of the source lifetime is negligible. $\beta_{\rm Tmax}$ are set as 0.1, 0.2, 0.3, 0.4, 0.5, 0.6, then the $\eta_{\rm Tmax}$ values are calculated by $\eta_{\rm Tmax}=\frac{1}{2}{\rm ln}\left( \frac{1+\beta_{\rm Tmax}}{1-\beta_{\rm Tmax}} \right)$. 

\begin{figure}[hbt]
	\centering
	\includegraphics[scale=0.32]{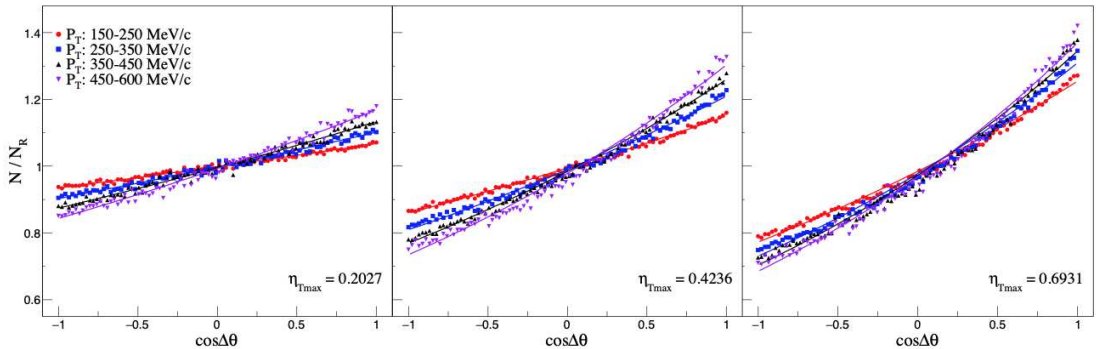}
	\caption{Fit normalized $\cos(\Delta\theta)$ for different $\eta_{\rm Tmax}$. The dots are normalized numbers of pions, and the lines are fit lines.}
	\label{cy_phifit}
\end{figure}

We generate pions which have random $\bm p_{\rm T}$ and random $\bm r_{\rm T}$, and use them to calculate the random $\cos(\Delta\theta)$ distribution. Then we use the distribution of $\cos(\Delta\theta)$ which is calculated from cylinder expansion source and divided by the random $\cos(\Delta\theta)$ distribution, to get the normalized $\cos(\Delta\theta)$ distribution. We fit normalized $\cos(\Delta\theta)$ distribution with equation~(\ref{consinfit}). The fit lines are shown in Figure~\ref{cy_phifit}, and the fit results are listed in Table 1.

\begin{table}[thb]
\label{fitresults}
\caption{Fit results of normalized $\cos(\Delta\theta)$ distribution.} 
\begin{indented}
\lineup
\item[]
\begin{tabular}{@{}*{6}{c}}
\br                              
$\eta_{\rm Tmax}$ & par & $p_{\rm T}:$150-250 MeV & 250-350 MeV & 350-450 MeV & 450-600 MeV\\
\mr
\multirow{2}{*}{0.1003} & $c_1$ & $0.9994\pm0.0002$ & $0.9990\pm0.0003$ & $0.9981\pm0.0003$ & $		0.9976\pm0.0003$ \\
{ } 					& $c_2$ & $0.0339\pm0.0003$ & $0.0496\pm0.0004$ & $0.0614\pm0.0004$ & $		0.0852\pm0.0005$ \\
\mr
\multirow{2}{*}{0.2027} & $c_1$ & $0.9985\pm0.0002$ & $0.9974\pm0.0003$ & $0.9954\pm0.0003$ & $		0.9924\pm0.0003$ \\
{ } 				  	& $c_2$ & $0.0663\pm0.0003$ & $0.0990\pm0.0004$ & $0.1299\pm0.0004$ & $		0.1621\pm0.0005$ \\
\mr
\multirow{2}{*}{0.3095} & $c_1$ & $0.9969\pm0.0002$ & $0.9937\pm0.0003$ & $0.9905\pm0.0003$ & $		0.9851\pm0.0003$ \\
{ } 				  	& $c_2$ & $0.1043\pm0.0003$ & $0.1485\pm0.0004$ & $0.1903\pm0.0004$ & $		0.2316\pm0.0005$ \\
\mr		
\multirow{2}{*}{0.4236} & $c_1$ & $0.9940\pm0.0002$ & $0.9887\pm0.0003$ & $0.9836\pm0.0003$ & $		0.9775\pm0.0003$ \\
{ } 				  	& $c_2$ & $0.1443\pm0.0004$ & $0.1990\pm0.0004$ & $0.2446\pm0.0004$ & $		0.2842\pm0.0005$ \\
\mr		
\multirow{2}{*}{0.5493} & $c_1$ & $0.9890\pm0.0002$ & $0.9829\pm0.0003$ & $0.9766\pm0.0003$ & $		0.9718\pm0.0003$ \\
{ } 				  	& $c_2$ & $0.1914\pm0.0004$ & $0.2467\pm0.0004$ & $0.2893\pm0.0008$ & $		0.3186\pm0.0005$ \\
\mr		
\multirow{2}{*}{0.6931} & $c_1$ & $0.9817\pm0.0003$ & $ 0.9735\pm0.0003 $ & $0.9683\pm0.0003$ & $		0.9631\pm0.0004$ \\
{ } 				  	& $c_2$ & $0.2450\pm0.0004$ & $ 0.2945\pm0.0004 $ & $0.3270\pm0.0005$ & $		0.3480\pm0.0006$ \\
\br
\end{tabular}
\end{indented}
\end{table}

\begin{figure}[htb]
	\centering
	\subfigure[]
	{
		\label{c1beta} 
		\includegraphics[width=0.40\textwidth]{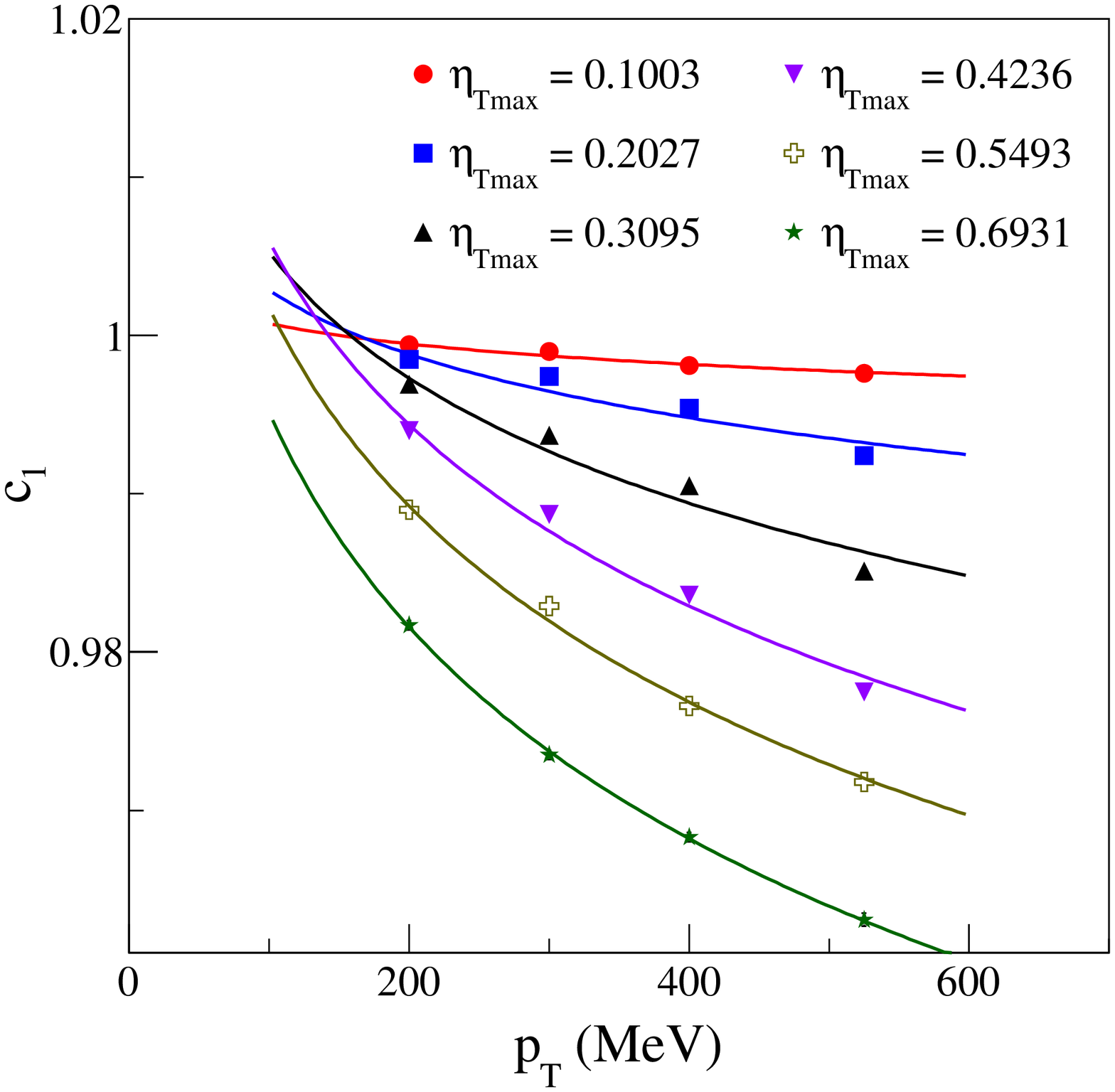}
	}
	\subfigure[]
	{
		\label{c2beta} 
		\includegraphics[width=0.40\textwidth]{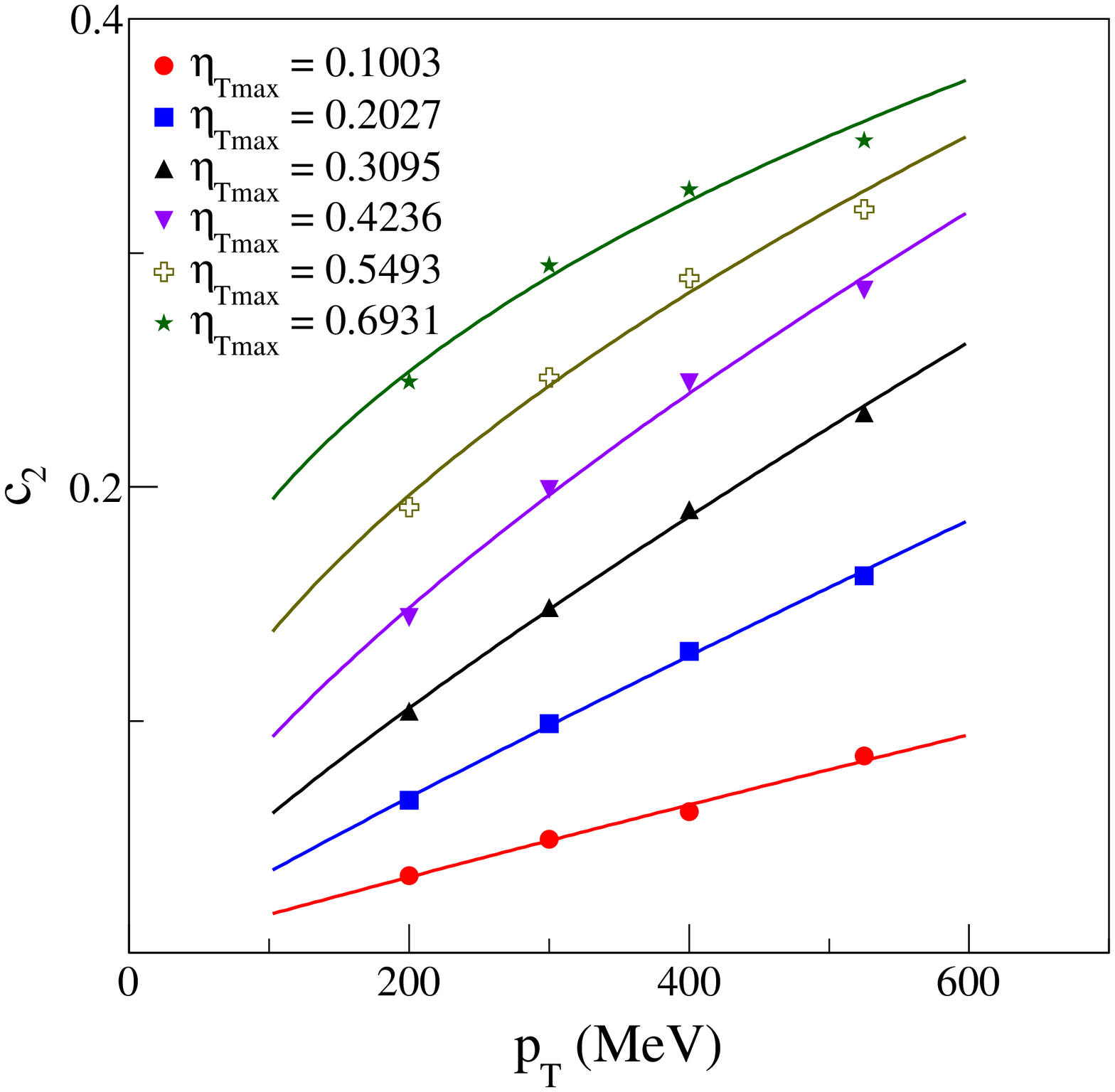}
	}
	
	\caption{Fit $c_1$ and $c_2$ for different $\eta_{\rm Tmax}$. The lines are fit lines.}
	\label{fit_c1_c2}
	~\\
\end{figure}

From the fit results, with increase of $p_{\rm T}$ and $\eta_{\rm Tmax}$, $c_1$ becomes smaller and $c_2$ becomes larger. We find $c_1$ and $c_2$ can be fitted by 
\begin{eqnarray}
	c_1= k_1p_{\rm T}^{j_1},\\
	c_2= k_2p_{\rm T}^{j_2},
\end{eqnarray}
where $k_1$, $k_2$ and $j_1$, $j_2$ are fit parameters. The fit lines are shown in Figure~\ref{fit_c1_c2}.

\begin{figure}[htb]
	\centering
	\subfigure[]
	{
		\label{HBTbeta=0.2} 
		\includegraphics[width=0.30\textwidth,natwidth=525,natheight=725]{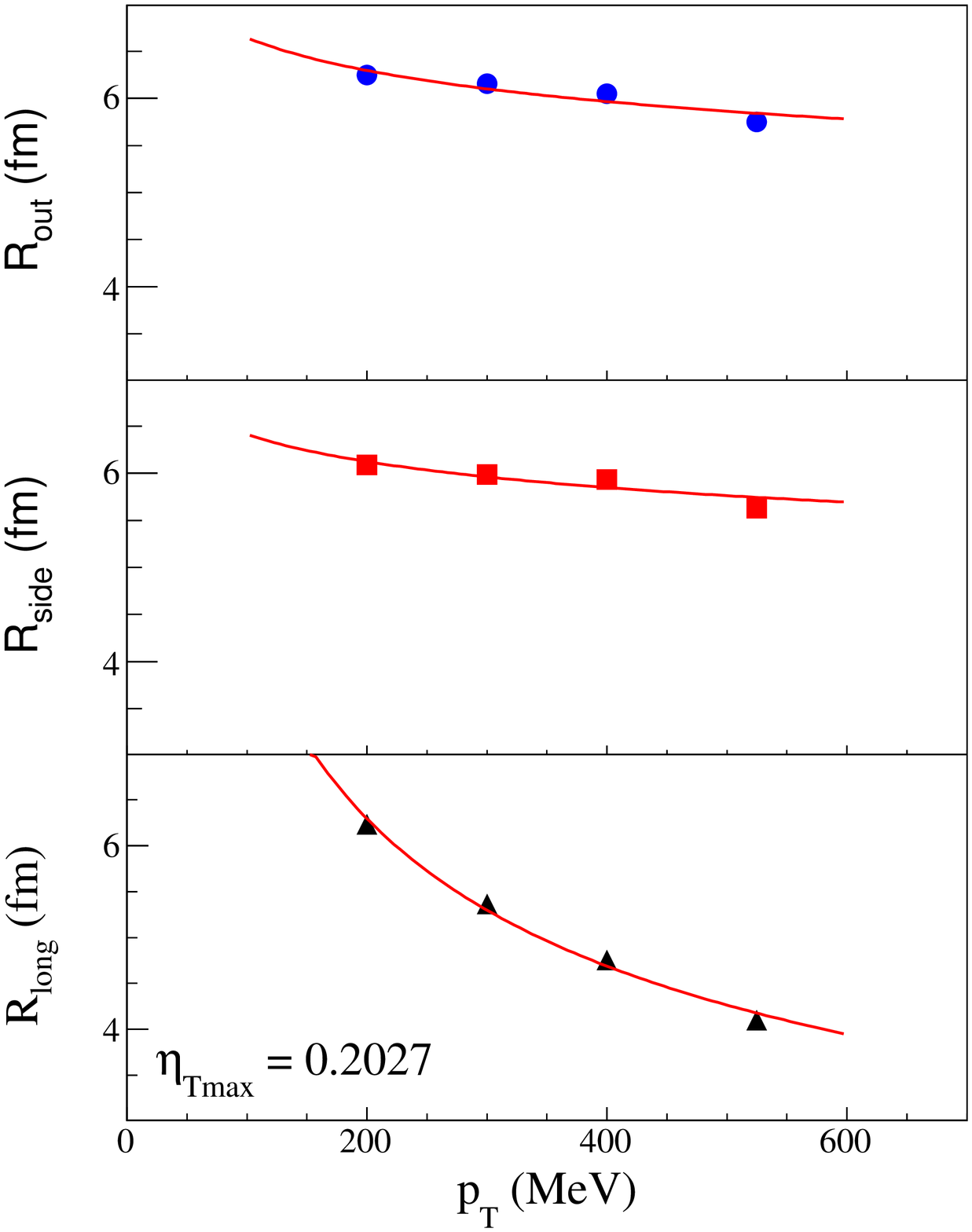}
	}
	\subfigure[]
	{
		\label{HBTbeta=0.4} 
		\includegraphics[width=0.30\textwidth,natwidth=525,natheight=725]{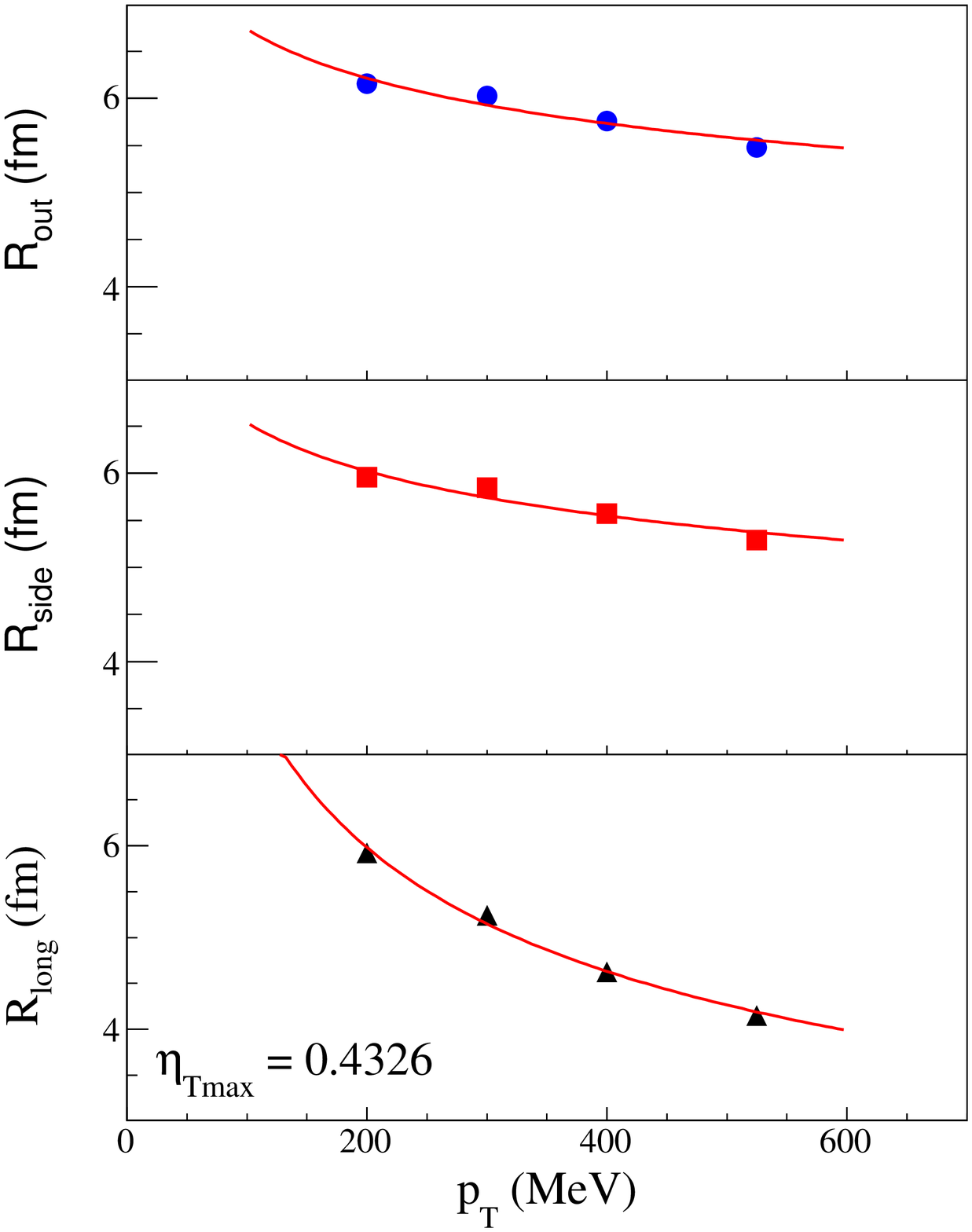}
	}
	\subfigure[]
	{
		\label{HBTbeta=0.6} 
		\includegraphics[width=0.30\textwidth,natwidth=525,natheight=725]{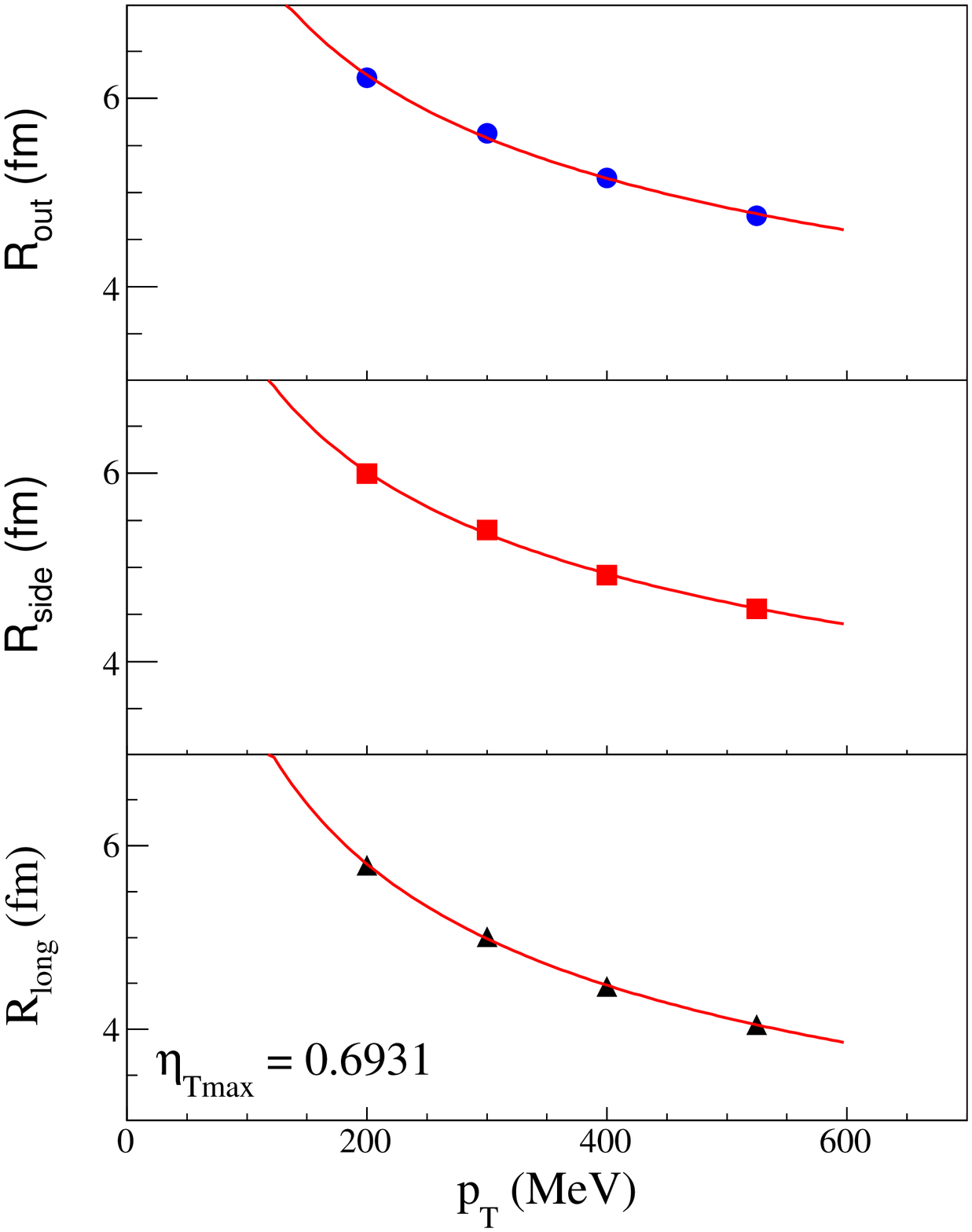}
	}
	\caption{ Fit HBT radii for different $\eta_{\rm Tmax}$. The red lines are fit lines.}
	\label{fit_HBT}
\end{figure}

We can also fit the HBT radii in different $p_{\rm T}$ regions by
\begin{equation} 
	R = a p_{\rm T}^{b},
\end{equation} 
where $a$ and $b$ are fit parameters. The fit lines are shown in Figure~\ref{fit_HBT}.

In the Figure~\ref{fit_HBT}, there are three cases of $\beta_{\rm Tmax} = 0.2, 0.4, 0.6$ for $\eta_{\rm Tmax} = 0.2027, 0.4236, 0.6931$. With the increase of $\eta_{\rm Tmax}$, the transverse flow becomes stronger, and the values of parameter b for $R_{\rm out}$ and $R_{\rm side}$ become larger. It indicates that the parameter $b$ describes the strength of $p_{\rm T}$ dependence, the lager $|b|$, the more prominent $p_{\rm T}$ dependence. So we only focus on the dependence of $\Delta \theta$ distribution on the parameter $b$.

\begin{figure}[!htb]
	\centering
	\subfigure[]
	{
		\label{b_k1}
		\includegraphics[width=0.45\textwidth,natwidth=700,natheight=725]
		{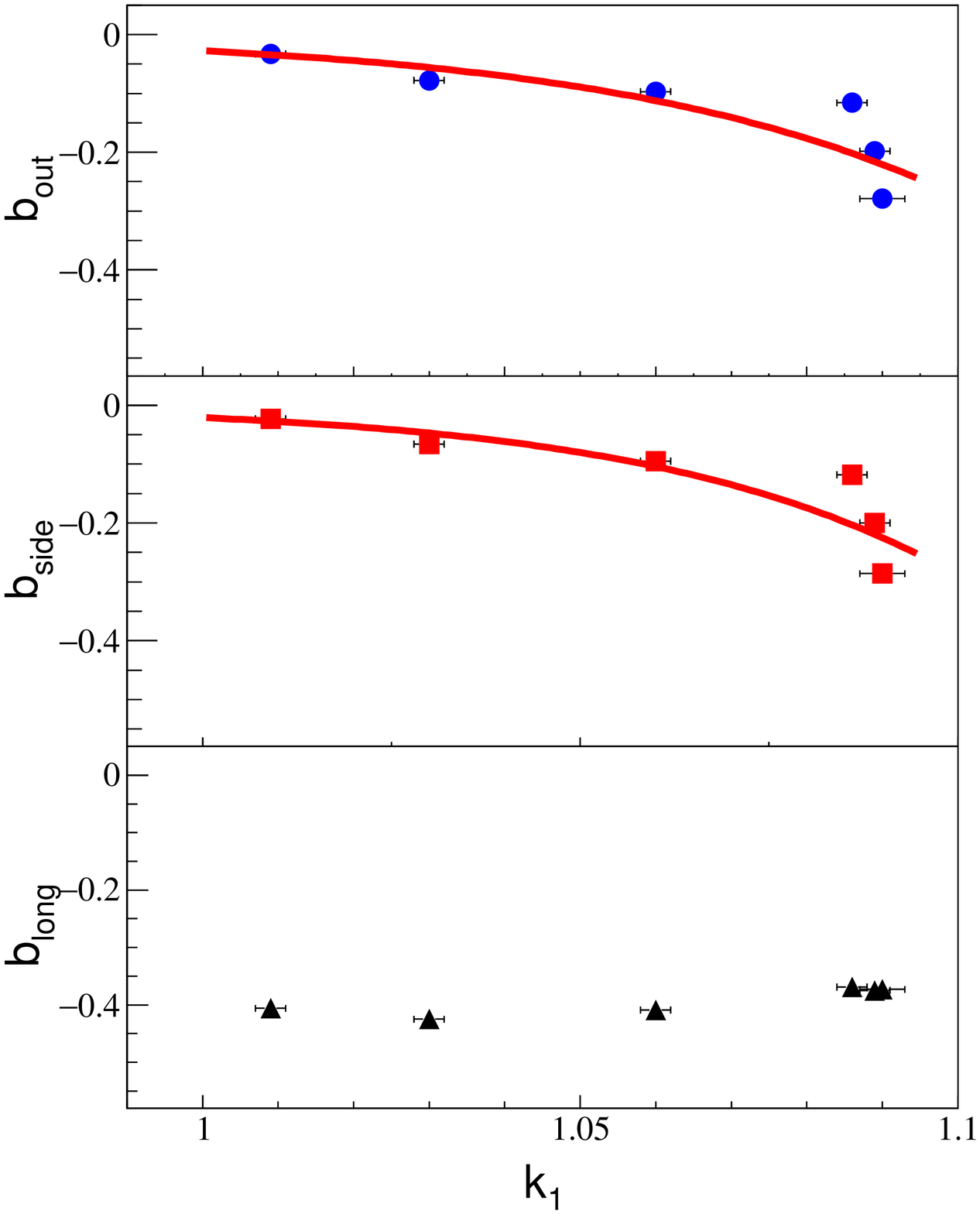}
	}
	\subfigure[]
	{
		\label{b_j1} 
		\includegraphics[width=0.45\textwidth,natwidth=700,natheight=725]
		{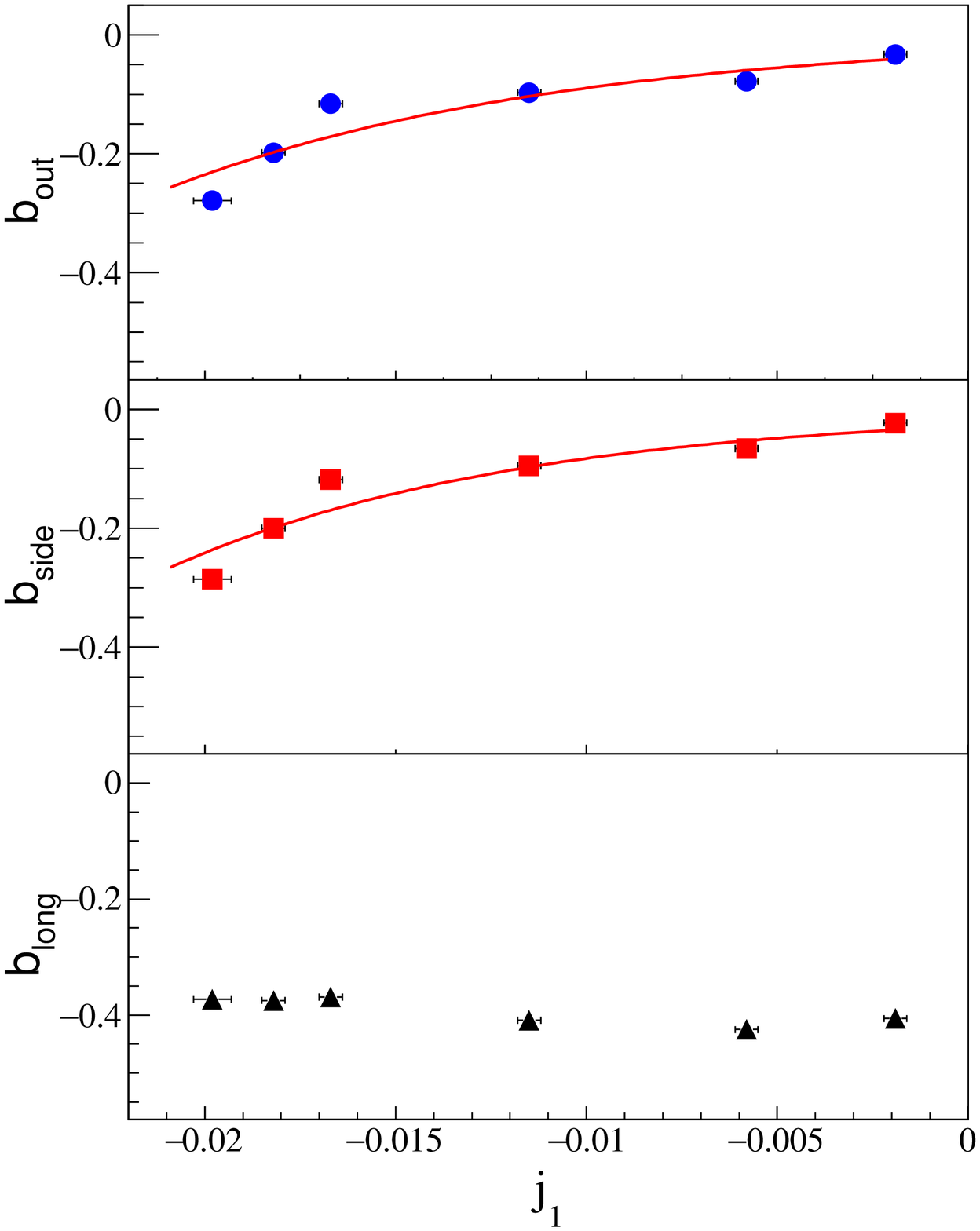}
	}
	
	\subfigure[]
	{
		\label{b_k2}
		\includegraphics[width=0.45\textwidth,natwidth=700,natheight=725]
		{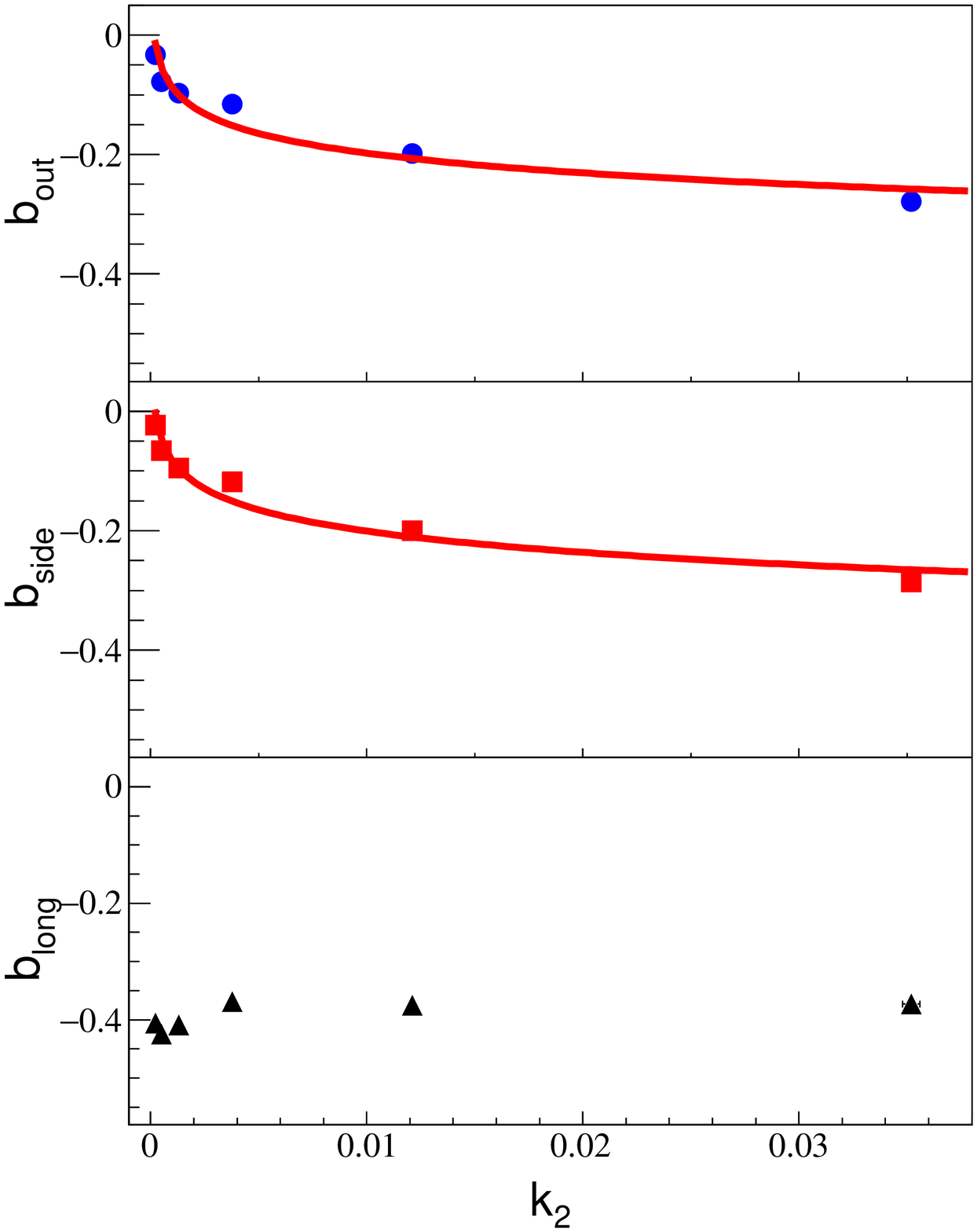}
	}
	\subfigure[]
	{
		\label{b_j2} 
		\includegraphics[width=0.45\textwidth,natwidth=700,natheight=725]
		{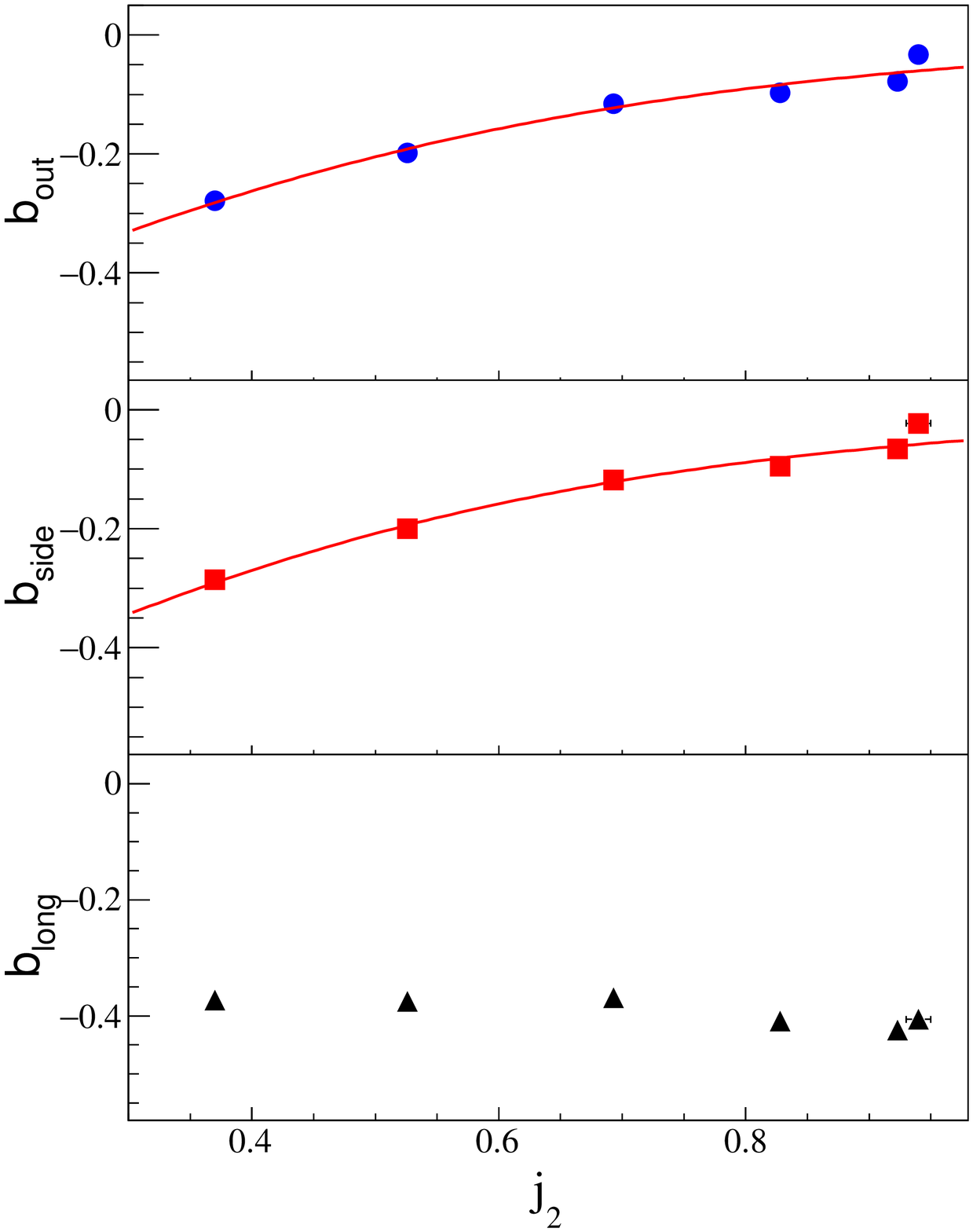}
	}
	\caption{Fit parameters in cylinder expansion source. Parameter $b$ from HBT radii fit function $R = a p_{\rm T}^{b}$, $j_1$, $j_2$ and $k_1$, $k_2$ are from fit function ${c_1}={k_1}p_{\rm T}^{j_1}$ and ${c_2}={k_2}p_{\rm T}^{j_2}$, $c_1$ and $c_2$ are from normalized space-momentum angle distribution function $f(\Delta\theta)=c_1{\rm{exp}}(c_2\cos(\Delta\theta))$, and the red lines are fit lines.}
	\label{b_j_k}
\end{figure}

We plot parameters $b$ as the function of $k$ and $j$ in figure~\ref{b_j_k}. Because of the longitudinal limit, there is barely any changes of $b_{long}$. The parameters in out and side directions are basically the same because the source lifetime is small enough. It indicates that there is connection between the HBT radii and the $\Delta \theta$ distribution. The red lines are fit lines and the fit functions are 
\begin{eqnarray}
	b(k_1) = \mu_{11} {k_1} ^{\mu_{12}},\label{bk1}                       \\
	b(j_1) = \nu_{11}{\rm e} ^{-\nu_{12} j_{1}},\label{bj1}    \\
	b(k_2) = \mu_{21} \ln{k_2}+\mu_{22},\label{bk2}                        \\
	b(j_2) = \nu_{21} \left(\frac{1}{1+\rm{e}^{\nu_{22} j_2}}-1\right).\label{bj2}
\end{eqnarray}
The fit parameters values are shown in Table~\ref{muandnu}. 
\begin{table}[!htb]
\begin{indented}
\lineup
\item[]
	\caption[table3]{Fit results of $b(k)$ and $b(j)$}\label{muandnu}
	\begin{tabular}{@{}*{4}{c}}
		\br
		\centre{2}{Parameters} & $b_{\rm {o}}$ & $b_{\rm {s}}$ \\
		\mr
		\multirow{4}{*}{$c_1$} 	& $\mu_{11}$ & $-0.026\pm0.001	$ & $-0.021\pm0.001	 $ 	\\
		{} 						& $\mu_{12}$ & $-24.3\pm0.5 	$ & $-28\pm1 		 $ 	\\
		{} 						& $\nu_{11}$ & $-0.034\pm0.002  $ & $-0.028\pm0.002	 $ 	\\
		{} 						& $\nu_{12}$ & $97\pm3    		$ & $107\pm4  		 $ 	\\
		\mr
		\multirow{4}{*}{$c_2$} 	& $\mu_{21}$ & $-0.0478\pm0.0006$ & $-0.0514\pm0.0006$ 	\\
		{} 						& $\mu_{22}$ & $-0.418\pm0.004  $ & $-0.437\pm0.004  $ 	\\
		{} 						& $\nu_{21}$ & $1.17\pm0.02     $ & $1.25\pm0.02  	 $ 	\\
		{} 						& $\nu_{22}$ & $-3.09\pm0.03  	$ & $-3.21\pm0.04	 $ 	\\
		\br
	\end{tabular}
\end{indented}
\end{table}

A connection has been made between the $p_{\rm T}$ dependence of HBT radii and the $\cos(\Delta\theta)$ distribution by equation~(\ref{bk1})-(\ref{bj2}) and Table~\ref{muandnu}, in a cylinder source. If all the pions freeze out almost at the same time, when we get a series of data of $R_{\rm out}$ or $R_{\rm side}$ in different $p_{\rm T}$ regions, we can describe the space-momentum angle distribution as a function of $p_{\rm T}$ in transverse plane. And if the space-momentum angle distribution is theoretically given as a function of $p_{\rm T}$ and the $R_{\rm out}$ or $R_{\rm side}$ is determined for a given $p_{\rm T}$ region, we can calculate $R_{\rm out}$ or $R_{\rm side}$ in other $p_{\rm T}$ regions. Because we limit the lifetime of source, the eight parameters of $b_{\rm o}$ are similar to the parameters of $b_{\rm s}$. If the source lifetime increases, $R_{\rm o}$ will also increase, and the parameters of $b_{\rm o}$ are no longer valid. And if we change the model, the value of parameters will also change.

\section{Conclusions}

By using several source models, we analyze the effect of source lifetime and single-particle space-momentum angle distribution on HBT radii. In the mid-rapidity region, $R_{\rm o}$ is sensitive to the source lifetime. $R_{\rm o}$ increases rapidly with the $\Delta t$. Furthermore, the HBT radii are also sensitive to the single-particle space-momentum angle distribution. With decreasing single-particle space-momentum angle, the HBT radii will also decrease. The collective expansion of the source leads to the changes of the single-particle space-momentum angle distribution with different $p_{\rm T}$, then causes the changes of HBT radii, at last, creates the transverse momentum dependence of HBT radii. In transverse plane of a cylinder expansion source, a numerical connection between the transverse momentum dependence of HBT radii and the the single-particle space-momentum angle distribution has been created. The parameters will change with different sources. If the parameters are settled, we can describe the transverse momentum dependence of HBT radii by the single-particle space-momentum angle distribution.

\section{Acknowledgments}
We appreciate the help of Miaomiao An for discussing the details of this work. We would also like to thank Xiaoze Tan and Weicheng Huang for valuable advice of this paper. 
\section*{References}
\bibliography{references}

\end{document}